%% file: Xi0Xi0.tex
\documentclass[prd,aps,reprint,noeprint,superscriptaddress,amsmath, amssymb]{revtex4-1}
\usepackage{graphicx}
\usepackage{subfigure}
\usepackage{epsfig}
\usepackage{multirow}
\usepackage{overpic}
\usepackage{color}
\usepackage{bm} 
\usepackage{xspace} 
\usepackage{dcolumn}
\usepackage{booktabs,tabularx}
\usepackage{array}
\usepackage{textcomp}
\usepackage[colorlinks,linkcolor=blue,urlcolor=blue,citecolor=blue]{hyperref}
\usepackage{tikz}

\usepackage{cleveref}

\newcommand{\jpsi}{J/\psi}

\begin{document}
\normalsize
\parskip=5pt plus 1pt minus 1pt


\title{\boldmath 
Tests of $CP$ symmetry in the entangled  $\Xi^0-\bar{\Xi}^0$ pairs }

\author{\input{authorlist_2023-02-20.tex}}
\begin{abstract}
    The $J/\psi \to \Xi^0 \bar{\Xi}^{0}$ process and subsequent decays
    are investigated using $(10087 \pm 44)\times 10^6$ $J/\psi$ events
    collected at the BESIII experiment. The decay parameters of
    $\Xi^0$ and $\bar{\Xi}^0$ are simultaneously measured to be $\alpha_{\Xi} = -0.3750
    \pm 0.0034 \pm 0.0016$, $\bar{\alpha}_{\Xi} = 0.3790 \pm 0.0034
    \pm 0.0021$, $\phi_{\Xi} = 0.0051 \pm 0.0096 \pm 0.0018$~rad,
    $\bar{\phi}_{\Xi} = -0.0053 \pm 0.0097 \pm 0.0019$~rad with unprecedented accuracies, where the
    first and the second uncertainties are statistical and systematic,
    respectively. The most precise values for $CP$ asymmetry observables of
    $\Xi^0$ decay are obtained to be $A^{\Xi}_{CP} = (-5.4
    \pm 6.5 \pm 3.1) \times 10^{-3}$ and $\Delta\phi^{\Xi}_{CP} =
    (-0.1 \pm 6.9 \pm 0.9) \times 10^{-3}$~rad. 
    For the first time, the weak and strong phase differences are determined
    to be $\xi_{P}-\xi_{S} = (0.0 \pm 1.7 \pm 0.2) \times
    10^{-2}$~rad and $\delta_{P}-\delta_{S} = (-1.3 \pm 1.7 \pm 0.4)
    \times 10^{-2}$~rad, which are the most precise results for any weakly-decaying baryon. 
    These results will play important roles in the studies of the $CP$ violations and polarizations for the strange, charmed and beauty baryons. 

\end{abstract}

\maketitle

At present, there is no satisfactory explanation for why our universe
is matter dominated. 
Following Sakharov~\cite{Sakharov:1967dj}, generation of a
matter-antimatter imbalance requires the fulfillment of three
criteria.  One of these is the existence of processes that violate
charge conjugation and parity ($CP$).  $CP$ violation ($CPV$) is
accommodated in the Standard Model of particle physics through the
Cabbibo-Kobayashi-Maskawa (CKM) mechanism and is experimentally
established in the meson sector~\cite{Christenson:1964fg,
  Aubert:2001nu, Abe:2001xe, Aaij:2019kcg}. However, the observed
$CPV$ in meson decays can only generate a matter-antimatter asymmetry
that is eight orders of magnitude smaller than that in our
universe~\cite{Bernreuther:2002uj, Canetti:2012zc}. The advent of high-intensity facilities producing hyperons and
antihyperons in abundance, opens up a new possibility: the search for
$CPV$ in hyperon decays~\cite{Donoghue:1985ww, Donoghue:1986hh,
  Salone:2022lpt, Li:2016tlt}.  

Hyperon decays are valuable since they can provide a way to measure $\Delta S = 1$ $CP$ nonconservation, 
which is complementary to kaon decays in which the $\Delta S = 2$ effects are dominant.
The observed $CPV$ from the $\Delta S = 2$ contributions in kaons decays can be well described by the CKM mechanism.
However, in CKM mechanism $\Delta S = 1$ effects also can be produced through a penguin diagram~\cite{Shifman:1975tn, Gilman:1980di}, 
this generates a nonzero value of the kaon decay parameter $\epsilon'$.
The penguin diagram could produce $CP$-odd effects at the order $20\epsilon'$ in hyperon decays, therefore hyperons have more potential to discover these effects.
The Weinberg-Higgs model~\cite{Weinberg:1976hu, Lee:1973iz} and leftright-symmetric model~\cite{Mohapatra:1974hk, Branco:1982wp, Beall:1981zq} also predict the $\Delta S = 1$ $CPV$.
Therefore, it is crucial to search for $CPV$ in hyperon decays,
and it has been demonstrated in several
measurements by the BESIII collaboration, where analyses of polarized
and entangled pairs of single-strange $\Lambda$~\cite{BESIII:2018cnd,
  BESIII:2022qax} and $\Sigma^+$~\cite{BESIII:2020fqg} hyperons
resulted in the most precise $CP$ tests so far for baryons.
Sequential decays of double-strange $\Xi$ hyperons are more intriguing, since they allow the separation
of strong and weak phase differences, as demonstrated by the
measurements of $\Xi^-$~\cite{BESIII:2021ypr, BESIII:2022lsz}.  
Its isospin partner, {\it i.e.} the $\Xi^0$ hyperon, provides independent
measurements of the $s \to u$ transition and the weak and strong phase
differences. 
In many previous experiments, the $CP$ tests rely on the products of weak and strong phase differences. 
In this way, it is difficult to distinguish the contributions of weak interactions from that of strong interactions. 
In $\Xi$ hyperon sequential decays, the separation of weak and strong phase differences allows us to directly determine 
the $CPV$ sources from Standard Model or it's beyond~\cite{Donoghue:1986hh, He:2022bbs}.

The decay amplitude of a spin-1/2 hyperon into a lighter spin-1/2
baryon and a pseudoscalar meson has a parity-violating S-wave
component and a parity-conserving P-wave component. Hence, it can be
completely described by the two independent decay parameters $\alpha$
and $\phi$~\cite{Lee:1957qs, Donoghue:1985ww, Donoghue:1986hh}.  The
decay parameters, $\alpha_{\Xi}(\alpha_{\Lambda}), \phi_{\Xi}, \bar{\alpha}_{\Xi}(\bar{\alpha}_{\Lambda}),
\bar{\phi}_{\Xi}$, of $\Xi^{0} (\Lambda$), and $\bar{\Xi}^{0} (\bar{\Lambda}$) 
hyperons can be determined from the sequential decays $\Xi^{0} \to \Lambda (\to p \pi^-)
\pi^{0}$ and $\bar{\Xi}^{0} \to \bar{\Lambda} (\to \bar{p} \pi^+) \pi^{0}$.
Precise measurements of the $\Xi^0$ decay parameters are 
important for studies of spin polarization and decay parameters of many other baryons
($\Omega^-, \Lambda_c, \Xi_c, \Xi_b, etc.$) decays into final states involving $\Xi^0$~\cite{BGHORS:1984jku, FOCUS:2005vxq, Han:2019axh}.

In this letter, we present measurements of $\Xi^0$ and $\bar{\Xi}^0$ decay
parameters and $CP$ asymmetries with a nine-dimensional fit to the
full angular distributions of the quantum-entangled $\Xi^0 -
\bar{\Xi}^0$ sequential decays.  

Three $CP$ asymmetry observables $A^{\Xi}_{CP}$, $\Delta \phi^{\Xi}_{CP}$,
and $A^{\Lambda}_{CP}$ are defined by the following equations:

\vspace{-0.5cm}
\begin{align}
    \label{eq:CPV}
    A^{\Xi}_{CP} &= (\alpha_{\Xi}+\bar{\alpha}_{\Xi})/(\alpha_{\Xi}-\bar{\alpha}_{\Xi}), \\
    \Delta\phi^{\Xi}_{CP} &= (\phi_{\Xi}+\bar{\phi}_{\Xi})/2, \\
    A^{\Lambda}_{CP} &= (\alpha_{\Lambda}+\bar{\alpha}_{\Lambda})/(\alpha_{\Lambda}-\bar{\alpha}_{\Lambda}),
\end{align}
since $CP$ conservation implies $\alpha_{\Xi}=-\bar{\alpha}_{\Xi}$,
$\phi_{\Xi}=-\bar{\phi}_{\Xi}$, and
$\alpha_{\Lambda}=-\bar{\alpha}_{\Lambda}$.  
According to Ref.~\cite{Salone:2022lpt}, $A^{\Xi}_{CP}$ is
proportional to the product of the weak phase difference $(\xi_P - \xi_S)$ and 
the strong phase difference $(\delta_P - \delta_S)$ of the final state
interaction.  Hence, for the case of a tiny strong phase
difference, $A^{\Xi}_{CP}$ would vanish even if the weak phase
difference is non-zero.  However, $\Delta\phi^{\Xi}_{CP}$ does not
have this problem~\cite{Salone:2022lpt} and is more sensitive than
$A^{\Xi}_{CP}$ for detecting $CPV$.  The weak and strong phase
differences can be determined from~\cite{Donoghue:1986hh, He:2022xra}

\vspace{-0.5cm}
\begin{align}
\label{eql:weakphase}
    \tan(\xi_{P}-\xi_{S})  = \frac{\sqrt{1-\alpha^2_{\Xi}}\sin\phi_{\Xi} + \sqrt{1-\bar{\alpha}^2_{\Xi}}\sin\bar{\phi}_{\Xi}}{\alpha_{\Xi} - \bar{\alpha}_{\Xi}},\\
    \tan(\delta_{P}-\delta_{S}) = \frac{\sqrt{1-\alpha^2_{\Xi}}\sin\phi_{\Xi} - \sqrt{1-\bar{\alpha}^2_{\Xi}}\sin\bar{\phi}_{\Xi}}{\alpha_{\Xi} - \bar{\alpha}_{\Xi}}.
\end{align}

This analysis is based on the sample of $(10087 \pm 44) \times 10^6$
$J/\psi$ events~\cite{BESIII:2021cxx} collected with the BESIII
detector at the BEPCII collider.  Details about BEPCII and BESIII can
be found in Refs. ~\cite{Yu:2016cof, Ablikim2010, BESIII:2020nme,
  Huang:2022wuo}.  A $J/\psi$ Monte Carlo (MC) simulation is used to
determine the detector efficiency, optimize the event selection, and
estimate the background.  The simulation is performed by {\sc
  geant4}-based~\cite{G42002iii} software~\cite{etde_20820412}, which
includes the geometric description of the BESIII detector and the
detector response.  The simulation models the beam energy spread and
initial state radiation (ISR) in the $e^+e^-$ annihilations with the
generator {\sc kkmc}~\cite{Jadach2001}.  The known decay modes of
$J/\psi$ are modeled with {\sc evtgen}~\cite{Lange2001,*Ping2008}, and
the remaining unknown decays are modeled with {\sc
  lundcharm}~\cite{Chen2000,*Yang2014a}.  For the signal process,
$\jpsi \to \Xi^{0} \bar{\Xi}^0, \Xi^{0} \to \Lambda (\to p \pi^-)
\pi^0,\bar{\Xi}^0 \to \bar{\Lambda} (\to \bar{p}\pi^+) \pi^0, \pi^0
\to \gamma \gamma$, two different MC samples are used.  One is
generated according to a phase space model (PHSP MC) for determination
of the parameters, and the other is generated according to the joint
angular distribution with the parameters obtained by this analysis
(signal MC).

The $\Lambda$ and $\bar{\Lambda}$ hyperons are reconstructed from their
dominant hadronic decay mode, $\Lambda(\bar{\Lambda}) \to p \pi^-
(\bar{p} \pi^+)$.  The charged tracks are detected in the Multilayer
Drift Chamber (MDC) under the requirement that $|\cos\theta|<0.93$,
where $\theta$ is the angle between the momentum of the
charged track and the axis of the detector.  Events with at least four
charged tracks are retained. Tracks with momentum larger than 0.3
GeV/$c$ are considered as proton candidates, otherwise as pion
candidates.  There are no further particle identification
requirements. Vertex fits~\cite{Xu:2009zzg} are performed using
all combinations with oppositely-charged proton and pion
candidates, constraining them to a common vertex. The combinations
which pass the vertex fit and have invariant masses within the
range $[1.111, 1.120]$~GeV/$c^2$ are regarded as $\Lambda$ and
$\bar{\Lambda}$ candidates.

The photons used for reconstructing the $\pi^0$ candidates are
detected in the electromagnetic calorimeter (EMC). Each photon is
required to have an EMC energy deposit of more than 25~MeV in the
barrel region ($|\cos\theta| < 0.80$) or more than 50~MeV in the
end-cap region ($0.86<|\cos\theta|<0.92$).  To suppress electronic
noise and showers unrelated to the event, the difference between the
EMC time and the event start time is required to be within (0,
700)~ns.  The $\pi^0$ candidates
require at least one photon to come from the barrel region and
invariant mass within the range $[0.098, 0.165]$ GeV/$c^2$.  A
kinematic fit~\cite{Yan:2010zze} constraining the $\gamma \gamma$
invariant mass to the known $\pi^0$
mass~\cite{ParticleDataGroup:2022pth} is performed, and the resulting
$\chi^2$ must be smaller than 200.

\begin{figure}[htbp]
    \begin{center}
        \mbox{
            \put(-125, 0){
                \begin{overpic}[width = 0.95\linewidth]{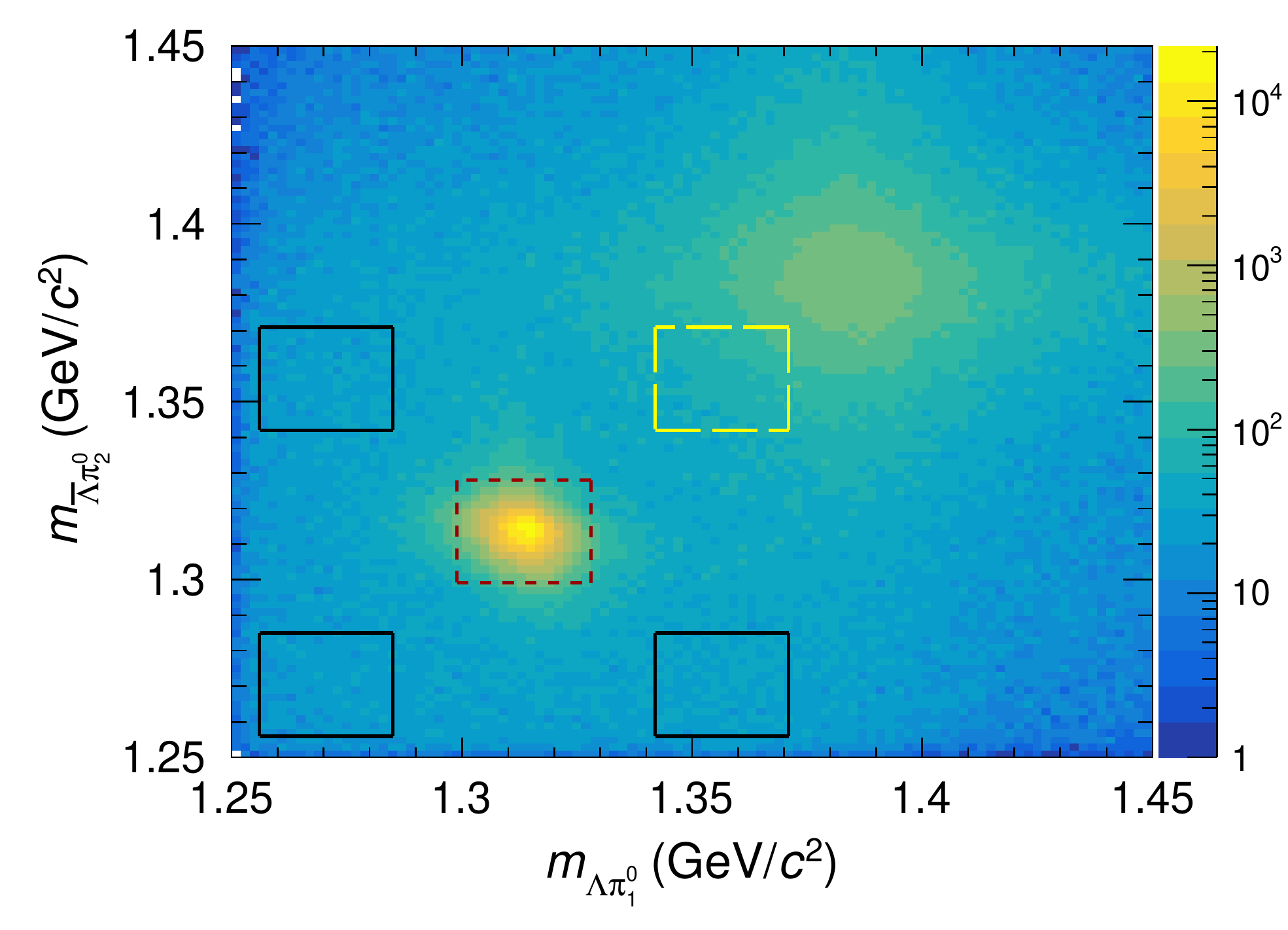}
                \end{overpic}
            }
        }
    \end{center}
    \caption{The distribution of $m_{\Lambda \pi^0_1}$ versus
      $m_{\bar{\Lambda} \pi^0_2}$, where the red short-dashed line box
      indicates the signal region, the black  and yellow
      long-dashed line boxes are the side-band regions.  }
    \label{mXi0vsmXi0bar}
\end{figure}

Events with at least one $\Lambda$ candidate, one $\bar{\Lambda}$
candidate, and two $\pi^0$ candidates are considered for further
analysis.  A four-constraint (4C) kinematic fit is performed to the
$\Lambda \bar{\Lambda} \pi^0 \pi^0$ hypothesis, constraining the total
reconstructed four-momentum to that of the initial $J/\psi$.  If there
is more than one combination of $\Lambda \bar{\Lambda} \pi^0 \pi^0$,
the one with the smallest $\chi^2$ of the 4C fit ($\chi^2_{4C}$) is
selected, and $\chi^2_{4C}< 100$ is required.  Since there are two
$\pi^0$ candidates ($\pi^0_1$, $\pi^0_2$) per event, there will be two
possible combinations of $\Lambda(\bar{\Lambda})$ and $\pi^0$. The
combination which minimizes the quantity $(m_{\Lambda \pi^0_1} -
M_{\Xi^0})^2 + (m_{\bar{\Lambda} \pi^0_2} - M_{\Xi^0})^2$ is kept,
where $m_{\Lambda \pi^0_1}$ and $m_{\bar{\Lambda} \pi^0_2}$ are the
invariant masses of $\Lambda \pi^0_1$ and $\bar{\Lambda} \pi^0_2$,
respectively, and $M_{\Xi^0}$ is the known mass of $\Xi^0$
~\cite{ParticleDataGroup:2022pth}.  The requirements suppress the
amount of miscombinations to approximately 0.7\%.  To reduce
background from sources with the same final states, {\it e.g.}
$J/\psi \to \Sigma^0(1385)\bar{\Sigma}^0(1385), \Sigma^0(1385)\to
\Lambda(\to p\pi^-) \pi^0, \bar{\Sigma}^0(1385) \to \bar{\Lambda}(\to
\bar{p} \pi^+) \pi^0$, $\pi^0 \to \gamma \gamma$, the $m_{\Lambda
  \pi^0_1}$ and $m_{\bar{\Lambda} \pi^0_2}$ are required to be within
$[1.299, 1.328]$ GeV/$c^2$. Figure~\ref{mXi0vsmXi0bar} shows the two-dimensional distribution of the
reconstructed $\bar{\Xi}^0$ mass versus the reconstructed $\Xi^0$ mass. 
After applying all the event selection criteria, we
obtain a sample of 327305 events.

An inclusive MC sample of 10 billion $J/\psi$ events is used for
studying potential backgrounds.  After applying the same selection
criteria as for data, the main background contribution is found to be
from the decay $J/\psi \to \Sigma^0(1385) \bar{\Sigma}^0(1385)$.  A
dedicated simulation of the process $J/\psi \to \Sigma^0(1385)
\bar{\Sigma}^0(1385)$ according to the measured angular distribution by Ref.~\cite{Ablikim:2016sjb} is carried out, and the corresponding number of
background events from this channel in data is estimated to
be $1697\pm139$.  All other background contributions are estimated
from the side-band regions of the two-dimensional distribution of
$m_{\Lambda \pi^0_1}$ versus $m_{\bar{\Lambda} \pi^0_2}$.  The
side-band regions are defined by $|m_{\Lambda
  \pi^0}(m_{\bar{\Lambda}\pi^0}) - M_{\Xi^0}| \in
[0.0285, 0.0575]$ GeV/$c^2$ and shown as black and yellow
long-dashed line boxes in Fig.~\ref{mXi0vsmXi0bar}. The events in the
yellow side-band box are mainly from $\Sigma^0(1385)$ and are excluded when
evaluating the other background contributions.  Since there are also
contributions from $\Sigma^0(1385)$ in the black box regions, the number
of background events other than $\Sigma^0(1385)$ is estimated by
$(N^{data}_{black} - N^{\Sigma^0(1385)}_{black})/3$, where
$N^{data}_{black}$ and $N^{\Sigma^0(1385)}_{black}$ are the numbers of
events in the black boxes from data and the MC simulated $\Sigma^0(1385)$
sample, respectively.  Finally, the number of background events other than
$\Sigma^0(1385)$ in the signal region is estimated to be $4641\pm138$.
The final selected data sample has a high signal purity of $(98.1\pm
0.2)$\%.

\begin{figure}[htbp]
    \begin{center}
        \begin{tikzpicture}[scale=1.0]
            \node(a) at (-1.0,0.0)
            {\includegraphics[width=1.0\linewidth]{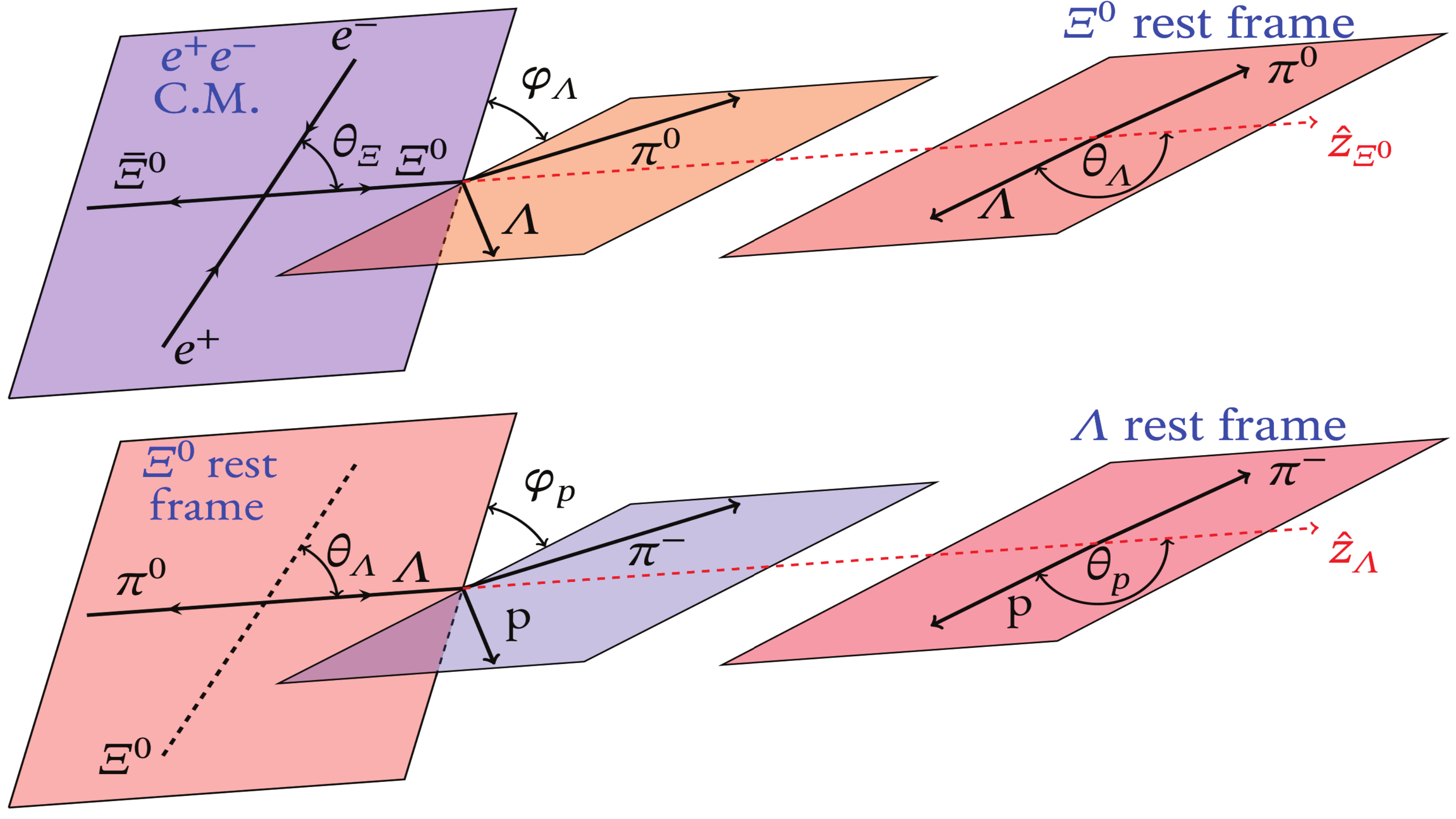}};
        \end{tikzpicture}
        \caption{The definitions of the helicity angles. The polar
          angle $\theta_{\Xi}$ is the angle between $\Xi^0$ momentum
          and $e^+$ beam direction in the $e^+ e^-$ center-of-mass
          system (C.M.), where the $\hat{z}$ axis is defined along the
          $e^+$ momentum.  $\theta_{\Lambda}$ and $\varphi_{\Lambda}$
          are the polar and azimuthal angles of the $\Lambda$ momentum
          direction in the $\Xi^0$ rest frame, where $\hat{z}_{\Xi^0}$
          is defined along the $\Xi^0$ momentum direction in the $e^+
          e^-$ C.M., and $\hat{y}_{\Xi^0}$ is defined by $\hat{z}
          \times \hat{z}_{\Xi^0}$.  The angles $\theta_{p}$ and
          $\varphi_{p}$ are the polar and azimuthal angles of the proton
          momentum direction in the $\Lambda$ rest frame, where
          $\hat{z}_{\Lambda}$ is defined along the $\Lambda$ momentum
          in the $\Xi^0$ rest frame, and $\hat{y}_{\Lambda}$ is
          along $\hat{z}_{\Xi^0} \times \hat{z}_{\Lambda}$.  }
        \label{DecayPlane}
    \end{center}
\end{figure}

Following the formulation in Refs.~\cite{Dubnickova:1992ii,
  Gakh:2005hh, Czyz:2007wi, Commins:1983ns}, the joint angular
distribution of the full decay chain is obtained, denoted as
$\mathcal{W}(\vec{\omega}; \vec{\zeta})$, and the final expression is
identical with the one developed by the helicity
frame~\cite{Perotti:2018wxm, BESIII:2021ypr}. Here, $\vec{\omega}$
represents the eight parameters of interest, $\alpha_{J/\psi}$,
$\Delta \Phi$, $\alpha_{\Xi}$, $\bar{\alpha}_{\Xi}$, $\phi_{\Xi}$,
$\bar{\phi}_{\Xi}$, $\alpha_{\Lambda}$, and $\bar{\alpha}_{\Lambda}$, where 
$\alpha_{J/\psi}$ and $\Delta\Phi$ are related to the psionic
form factors~\cite{Faldt:2017kgy} and govern the scattering angle
distribution and the polarization of the $\Xi^0$;
$\vec{\zeta}$ stands for nine angle variables, $\theta_{\Xi}$,
$\theta_{\Lambda}$, $\varphi_{\Lambda}$, $\theta_{\bar{\Lambda}}$,
$\varphi_{\bar{\Lambda}}$, $\theta_{p}$, $\varphi_{p}$,
$\theta_{\bar{p}}$, and $\varphi_{\bar{p}}$.  These
helicity angles, are constructed as illustrated in
Fig.~\ref{DecayPlane}, and the corresponding angles of the
anti-particle decay sequence are obtained analogously.  Detailed
information about this formalism can be found in
Ref.~\cite{BESIII:2021ypr}.

A 9-dimensional maximum likelihood fit is performed on the joint angular
distribution from data to determine the eight
$\vec{\omega}$ parameters, similar to that in
Ref.~\cite{BESIII:2022qax}. 
In the fit, the events from the data side-band regions and the $\Sigma^0(1385)$ MC sample are included with a negative weight to subtract the background effects, 
where the likelihood function of background events is the same as the signal events.
The results are
summarized in Table~\ref{tab:final_results}, together with previous
measurements~\cite{Ablikim:2016sjb, BESIII:2023lkg, BESIII:2022qax}.
In this table, we also present the averaged values of the decay parameters, which are defined as: $ \langle \alpha_{\Xi} \rangle = (\alpha_{\Xi} - \bar{\alpha}_{\Xi})/2$,
$\langle \phi_{\Xi} \rangle = (\phi_{\Xi} - \bar{\phi}_{\Xi})/2$,
$\langle \alpha_{\Lambda} \rangle = (\alpha_{\Lambda} - \bar{\alpha}_{\Lambda})/2$.
\begin{table}[htbp]
\renewcommand\arraystretch{1.2}
    \caption{The $J/\psi \to \Xi^0 \bar{\Xi}^0$ angular distribution parameter $\alpha_{J/\psi}$, the relative phase $\Delta \Phi$ of the psionic form factors, 
    the decay parameters for $\Xi^0 \to \Lambda \pi^0 (\alpha_{\Xi}, \phi_{\Xi})$, 
    $\bar{\Xi}^0 \to \bar{\Lambda} \pi^0 (\bar{\alpha}_{\Xi}, \bar{\phi}_{\Xi})$, 
    $\Lambda \to p \pi^+ (\alpha_{\Lambda})$ and $\bar{\Lambda} \to \bar{p} \pi^+ (\bar{\alpha}_{\Lambda})$;
    the $CP$ asymmetries $A^{\Xi}_{CP}$, $\Delta \phi^{\Xi}_{CP}$, and $A^{\Lambda}_{CP}$, and the averages $\langle \alpha_{\Xi} \rangle$,
    $\langle \phi_{\Xi} \rangle$, and $\langle \alpha_{\Lambda} \rangle$. The first and second uncertainties are
    statistical and systematic, respectively.
    } 
\label{tab:final_results}
\centering
\scalebox{0.9}
{
    \begin{tabular}{l c c c}
        \hline \hline 
        \noalign{\vskip 3pt}
        Parameter               & This work                                         & Previous result\\
        \hline
        $\alpha_{J/\psi}$       & ${\color{white}-}0.514  \pm 0.006 \pm  0.015$     & ${\color{white}-}0.66 \pm 0.06 $~\cite{Ablikim:2016sjb} \\
        $\Delta\Phi$(rad)            & ${\color{white}-}1.168  \pm 0.019 \pm  0.018$     & -                         \\
        $\alpha_{\Xi}$          & $-0.3750 \pm 0.0034 \pm 0.0016$                   & $-0.358\pm 0.044$~\cite{BESIII:2023lkg}        \\
        $\bar{\alpha}_{\Xi}$    & ${\color{white}-}0.3790 \pm 0.0034 \pm 0.0021$    & ${\color{white}-}0.363 \pm 0.043$~\cite{BESIII:2023lkg}                        \\
        $\phi_{\Xi}$(rad)            & ${\color{white}-}0.0051 \pm 0.0096 \pm 0.0018$    & ${\color{white}-}0.03 \pm 0.12$~\cite{BESIII:2023lkg}          \\
        $\bar{\phi}_{\Xi}$(rad)      & $-0.0053 \pm 0.0097 \pm 0.0019$                   & $-0.19\pm 0.13$~\cite{BESIII:2023lkg}\\
        $\alpha_{\Lambda}$      & ${\color{white}-}0.7551 \pm 0.0052 \pm 0.0023$    & ${\color{white}-}0.7519\pm 0.0043$~\cite{BESIII:2022qax}        \\
        $\bar{\alpha}_{\Lambda}$& $-0.7448 \pm 0.0052 \pm 0.0017$                   & $-0.7559\pm0.0047$~\cite{BESIII:2022qax}        \\
        \hline 
        $\xi_{P}-\xi_{S}$(rad)      & ${\color{white}-}(0.0  \pm 1.7  \pm 0.2)\times 10^{-2}$     & - \\ 
        $\delta_{P}-\delta_{S}$(rad) & $(-1.3  \pm 1.7  \pm 0.4)\times 10^{-2}$                    & - \\ 
        \hline 
        $A^{\Xi}_{CP}$          & $(-5.4 \pm 6.5 \pm 3.1)\times 10^{-3}$                      & $(-0.7\pm8.5)\times10^{-2}$~\cite{BESIII:2023lkg} \\  
        $\Delta\phi^{\Xi}_{CP}$(rad)      & $(-0.1 \pm 6.9 \pm 0.9)\times 10^{-3}$   & $(-7.9\pm8.3)\times10^{-2}$~\cite{BESIII:2023lkg} \\ 
        $A^{\Lambda}_{CP}$      & ${\color{white}-}(6.9 \pm 5.8 \pm 1.8) \times 10^{-3}$    & $(-2.5 \pm 4.8)\times 10^{-3}$~\cite{BESIII:2022qax} \\  
        \hline 
        $\langle \alpha_{\Xi} \rangle$        & $-0.3770 \pm 0.0024\pm 0.0014$                & - \\ 
        $\langle \phi_{\Xi} \rangle$(rad)          & ${\color{white}-}0.0052 \pm 0.0069 \pm 0.0016$& - \\ 
        $\langle \alpha_{\Lambda} \rangle$    & ${\color{white}-}0.7499 \pm 0.0029 \pm 0.0013$& ${\color{white}-} 0.7542 \pm 0.0026$~\cite{BESIII:2022qax}        \\
        \hline \hline
    \end{tabular}
    }
\end{table}

\begin{figure}[htbp]
    \begin{center}
        \mbox{
            \put(-125, 0){
                \begin{overpic}[width = 0.95\linewidth]{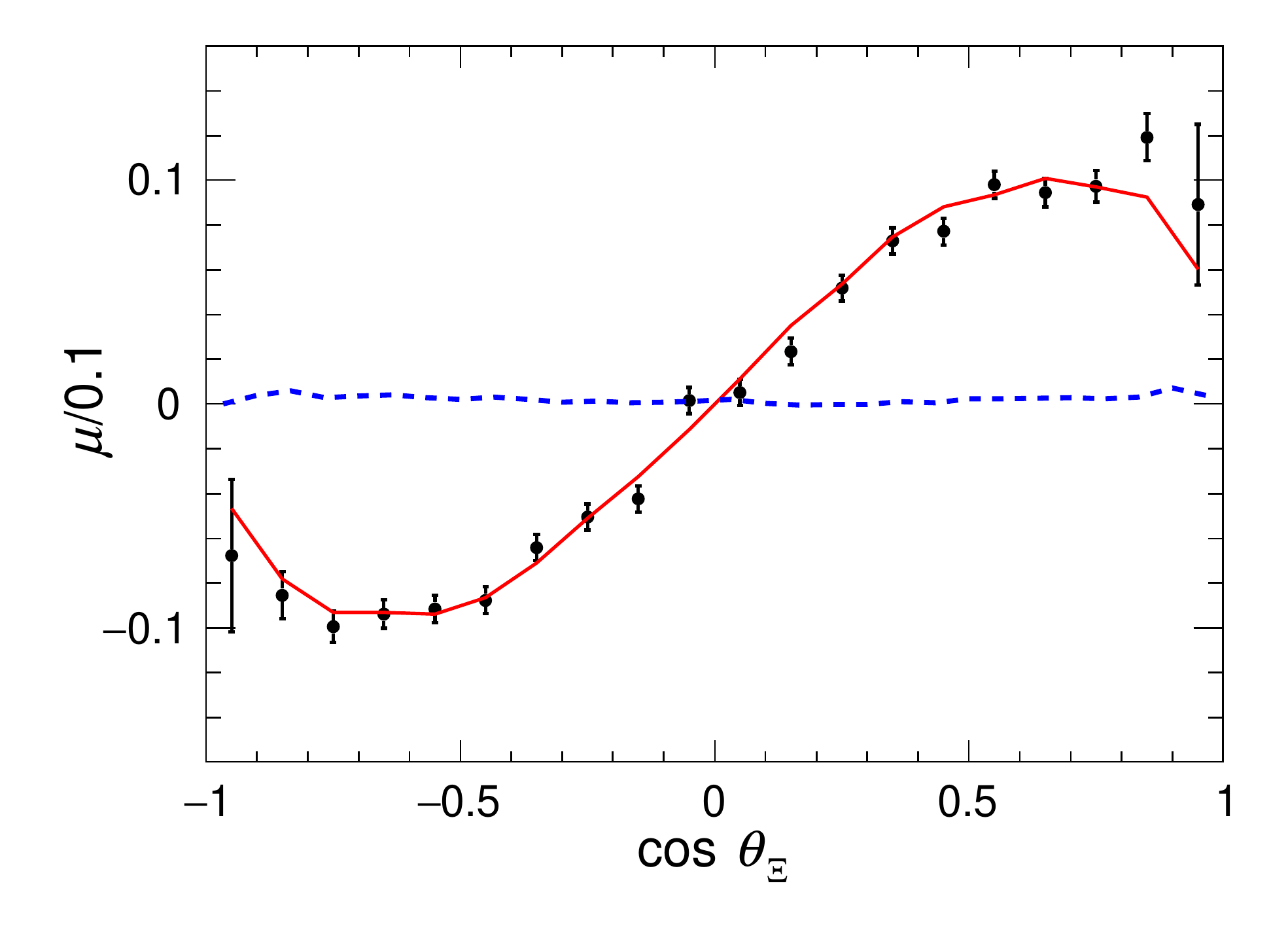}
                \end{overpic}
            }
        }
    \end{center}
    \caption{Distribution of the moment $\mu(\cos\theta_{\Xi})$ versus
      $\cos\theta_{\Xi}$. The points with error bars are data,
      the red solid lines are from the signal MC, and the blue dashed
      line represents the distribution without polarization from the PHSP
      MC sample.}
    \label{Polarization}
\end{figure}

The $\Xi^0$ polarization can be illustrated through the moment $\mu$, as defined: 

\vspace{-0.5cm}
\begin{align}
    \label{eq:mu}
    \mu^k(\cos\theta_{\Xi}) &= \frac{1}{N^{k}} \sum^{N^k}_{i} (\sin\theta^i_{\Lambda} \sin\varphi^i_{\Lambda} + \sin\theta^i_{\bar{\Lambda}} \sin\varphi^i_{\bar{\Lambda}}),
\end{align}
where $N^k$ is the number of events in k-th $\cos\theta_{\Xi}$ bin and $i$ is the $i$-th event in
that bin. The expected angular dependence of
the moment for the acceptance-corrected data is given by
\begin{equation}
    \mu(\cos\theta_{\Xi}) = \frac{\alpha_{\Xi} - \bar{\alpha}_{\Xi}}{2}\frac{1+\alpha_{J/\psi}\cos^2\theta_{\Xi}}{3+\alpha_{J/\psi}}P_{y}(\theta_{\Xi}),
\end{equation}
where $P_{y}(\theta_{\Xi}) = \sqrt{1-\alpha^{2}_{J/\psi}} \sin(\Delta
\Phi)\cos \theta_{\Xi}\sin \theta_{\Xi} /(1+\alpha_{J/\psi}\cos^2
\theta_{\Xi})$  is the polarization of $\Xi^0$.  Comparing
the data to the PHSP MC sample, as shown in Fig.~\ref{Polarization},
there is a significant polarization of the $\Xi^0$ hyperons produced in
$J/\psi \to \Xi^0 \bar{\Xi}^0$ decays, manifested in the relative phase 
$\Delta \Phi = 1.168\pm0.019_{\rm stat}\pm0.018_{\rm
  syst}$~radians.

\begin{table*}[!htbp]
\renewcommand\arraystretch{1.2}
    \caption{Absolute systematic uncertainties for the measured parameters.
    }
\label{tab:systematic_err}
\centering
\scalebox{0.86}
{
    \begin{tabular}{l c c c c c c c c | c c | c c c | c c c}
        \hline \hline 
        \noalign{\vskip 3pt}
        Source ($10^{-3}$)                      & $\alpha_{J/\psi}$ & $\Delta\Phi$(rad)  & $\alpha_{\Xi}$ & $\bar{\alpha}_{\Xi}$ & $\phi_{\Xi}$(rad) & $\bar{\phi}_{\Xi}$(rad) & $\alpha_{\Lambda}$ & $\bar{\alpha}_{\Lambda}$ & $\xi_{P}-\xi_{S}$(rad) & $\delta_{P}-\delta_{S}$(rad) & $A^{\Xi}_{CP}$ & $\Delta\phi^{\Xi}_{CP}$(rad) & $A^{\Lambda}_{CP}$ & $\langle \alpha_{\Xi} \rangle$ & $\langle \phi_{\Xi} \rangle$(rad) & $\langle \alpha_{\Lambda}\rangle$\\
        \hline
        $\Lambda/\bar{\Lambda}$ reconstruction  & 9 & 8 & 0.1 & 1.8 & 0.6 & 0.1 & 1.0 & 0.8 & 1 & 1 & 2.3 & 0.3 & 0.1 & 0.8 & 0.3 & 0.8  \\
        $\pi^0$ reconstruction                  & 2 & 6 & 0.5 & 0.4 & 1.2 & 1.1 & 0.0 & 0.0 & 0 & 3 & 0.1 & 0.1 & 0.0 & 0.5 & 1.1 & 0.1  \\
        4C Kinematic fit                        & 7 & 8 & 0.9 & 0.3 & 0.8 & 0.8 & 0.8 & 0.3 & 2 & 0 & 1.7 & 0.8 & 0.3 & 0.3 & 0.0 & 0.5 \\
        $\Xi^0/\bar{\Xi}^0$ mass window         & 3 & 7 & 0.5 & 0.0 & 0.0 & 0.0 & 0.0 & 0.0 & 0 & 0 & 0.0 & 0.0 & 0.0 & 0.4 & 0.0 & 0.0  \\
        Background estimation                   & 0 & 1 & 0.1 & 0.2 & 0.3 & 0.5 & 0.5 & 0.8 & 0 & 1 & 0.2 & 0.3 & 0.3 & 0.1 & 0.3 & 0.5  \\
        $\cos\theta_{\Xi}$ inconsistency        & 8 & 9 & 0.9 & 1.0 & 0.6 & 0.5 & 0.7 & 0.8 & 0 & 2 & 0.1 & 0.1 & 0.0 & 0.9 & 0.5 & 0.7  \\
        Fit method                              & 3 & 6 & 0.6 & 0.2 & 0.7 & 1.1 & 1.7 & 0.9 & 1 & 2 & 1.1 & 0.2 & 1.7 & 0.2 & 0.9 & 0.4  \\
        \hline                                                                                                   
        Total                                   &15 &18 & 1.6 & 2.1 & 1.8 & 1.9 & 2.3 & 1.7 & 2 & 4 & 3.1 & 0.9 & 1.8 & 1.4 & 1.6 & 1.3  \\ 
        \hline \hline
    \end{tabular}
    }
\end{table*}

The systematic uncertainties can be separated into two categories: 
(1) the differences between data and simulation,
(2) the uncertainties associated with the fit procedure. 

The systematic uncertainties from $\Lambda/\bar{\Lambda}$
reconstruction, $\pi^0$ reconstruction, and the kinematic fit are
studied with control samples. A control sample of $J/\psi \to
pK^-\bar{\Lambda} + c.c.$ is used to estimate the uncertainties from
the $\Lambda/\bar{\Lambda}$ reconstruction.  The systematic
uncertainties from $\pi^0$ reconstruction and 4C kinematic fit are
investigated by a control sample of $J/\psi \to \Xi^0(\to \Lambda \pi^0)
\bar{\Xi}^0 (\to \bar{\Lambda} \pi^0)$. The efficiency differences
between data and MC for the control samples are used to re-weight
the PHSP MC sample. The differences between the fitting results with
corrections and the nominal fitting are taken as the systematic
uncertainties.

The mass window of $\Xi^0/\bar{\Xi}^0$ is $\pm 3\sigma$ around the
known $\Xi^0$ mass, where $\sigma = 4.8$ MeV/$c^2$ is the resolution
of the reconstructed $\Xi^0/\bar{\Xi}^0$ mass. We change it to $\pm
2\sigma$ or $\pm 4\sigma$ to study the systematic uncertainties from
the $\Xi^0/\bar{\Xi}^0$ mass window. The fit is repeated, and the
largest deviations from the nominal values are taken as the signal
mass window systematic uncertainties.

The systematic uncertainties from the background estimation are
studied by varying the estimated background yields by one
standard deviation or changing the side-band regions from $|m_{\Lambda \pi^0}(m_{\bar{\Lambda}\pi^0}) - M_{\Xi^0}| \in [0.0285, 0.0575]$~GeV/$c^2$ to $|m_{\Lambda \pi^0}(m_{\bar{\Lambda}\pi^0}) - M_{\Xi^0}| \in [0.0235, 0.0525]$~GeV/$c^2$. 
The biggest differences between the results with the 
modified background yields and nominal ones are taken as the
systematic uncertainties.

On the left and right sides ($|\cos\theta_{\Xi}|>0.85$) of the $\cos\theta_{\Xi}$ distribution, 
the number of events in data is observed to be smaller than what in the MC simulations, where the total number of events of the MC sample is normalized to data.
To estimate their effect on the final
results, the ratios of $n^i_{data}/n^i_{MC}$ are obtained in different
$\cos\theta_{\Xi}$ bins, where $n^i_{data}$ and $n^i_{MC}$ are the
number of events in the $i$-th $\cos\theta_{\Xi}$ bin from the data
and MC sample, respectively. The ratios are then used to re-weight
the MC sample, and the differences between the results after
weighting and the nominal results are taken as systematic
uncertainties.

All the above systematic uncertainties belong to category (1); here we
consider category (2), the uncertainties caused by the fit method.
This source of systematic uncertainties is estimated by analyzing the
signal MC sample which is approximately one hundred times of our
nominal data sample. The differences between the obtained parameter
values and the ones we used to generate the signal MC sample are
regarded as systematic uncertainties.

The absolute systematic uncertainties for various sources are
summarized in Table~\ref{tab:systematic_err}. The total systematic
uncertainties of various parameters are obtained by summing the
individual contributions in quadrature.

In summary, this letter presents the most precise determination of all
$\Xi^0 \to \Lambda \pi^0/\bar{\Xi}^0 \to \bar{\Lambda} \pi^0$ decay
parameters, which are improved by more than one order of magnitude
over the previous measurements~\cite{BESIII:2023lkg}, as shown in
Table~\ref{tab:final_results}. 
The averaged values of the $\Xi^0$ decay parameters are determined to be 
$\langle \alpha_{\Xi} \rangle = -0.3770 \pm 0.0024_{\rm stat} \pm 0.0014_{\rm syst}$ 
and $\langle \phi_{\Xi} \rangle = 0.0052 \pm 0.0069_{\rm stat} \pm 0.0016_{\rm syst}$ for the first time, 
which will be valuable inputs for many other baryon studies involving $\Xi^0$ decay.
A clear transverse polarization of $\Xi^0$ from
$J/\psi$ decay is observed for the first time, the relative phase of the psionic form factors is determined to be $\Delta \Phi = 1.168 \pm 0.019_{\rm stat} \pm 0.018_{\rm syst}$~rad.
This result is significantly
different from the $\psi(3686) \to \Xi^0 \bar{\Xi}^0$ decay~\cite{BESIII:2023lkg}, $\Delta \Phi_{\psi(3686)} = -0.050 \pm 0.150_{\rm stat} \pm 0.020_{\rm syst}$, where no
polarization was found. These observations will provide a key probe of the decay dynamics on the charmonium decays to hyperon pairs.
The $CP$ asymmetry observables are measured to be 
$A^{\Xi}_{CP} = (-5.4\pm6.5_{\rm stat}\pm3.1_{\rm syst}) \times 10^{-3}$ and $\Delta\phi^{\Xi}_{CP} = (-0.1\pm6.9_{\rm stat}\pm0.9_{\rm syst}) \times 10^{-3}$
with the highest precision to date. We find $CP$ symmetry is conserved
in the $\Xi^0$ decay with an accuracy of $10^{-3}$, which is in
agreement with the Standard Model predictions~\cite{He:2022xra}.
Furthermore, the weak and strong phase differences in $\Xi^0$ decay,
$\xi_{P}-\xi_{S}= (0.0 \pm 1.7_{\rm stat} \pm 0.2_{\rm syst})\times 
10^{-2}$ and $\delta_{P}-\delta_{S} = (-1.3 \pm 1.7_{\rm stat} \pm 0.4_{\rm syst})\times 10^{-2}$, are directly measured for the first time. 
These are the most precise results for any weakly-decaying baryon and crucial for understanding $CP$ violation sources beyond the Standard Model.
In the same time, the decay parameters and $CP$ asymmetry observable of the decay $\Lambda \to p \pi^-/\bar{\Lambda} \to \bar{p} \pi^+$ 
are determined.
The parameter $\alpha_{\Lambda}$
is found to be in excellent agreement with the determinations from
$J/\psi \to \Lambda \bar{\Lambda}$ and $J/\psi \to \Xi^- \bar{\Xi}^+$
samples~\cite{BESIII:2021ypr, BESIII:2022qax}, but in 3.5$\sigma$
tension with the CLAS result, $0.721\pm 0.006_{\rm stat} \pm
0.005_{\rm syst}$~\cite{Ireland:2019uja}.
With the properties of
quantum-entangled hyperon-antihyperon, the proposed future super tau-charm
factories~\cite{Charm-TauFactory:2013cnj, Achasov:2023gey} may have  potential to discover $CPV$ in the baryon sector.

The BESIII Collaboration thanks the staff of BEPCII and the IHEP computing center for their strong support. This work is supported in part by National Key R\&D Program of China under Contracts Nos. 2020YFA0406300, 2020YFA0406400; National Natural Science Foundation of China (NSFC) under Contracts Nos. 11635010, 11735014, 11835012, 11935015, 11935016, 11935018, 11961141012, 12022510, 12025502, 12035009, 12035013, 12061131003, 12192260, 12192261, 12192262, 12192263, 12192264, 12192265, 12221005, 12225509, 12235017; the Chinese Academy of Sciences (CAS) Large-Scale Scientific Facility Program; the CAS Center for Excellence in Particle Physics (CCEPP); CAS Key Research Program of Frontier Sciences under Contracts Nos. QYZDJ-SSW-SLH003, QYZDJ-SSW-SLH040; 100 Talents Program of CAS; The Institute of Nuclear and Particle Physics (INPAC) and Shanghai Key Laboratory for Particle Physics and Cosmology; ERC under Contract No. 758462; European Union's Horizon 2020 research and innovation programme under Marie Sklodowska-Curie grant agreement under Contract No. 894790; German Research Foundation DFG under Contracts Nos. 443159800, 455635585, Collaborative Research Center CRC 1044, FOR5327, GRK 2149; Olle Engkvist Foundation, Contract No. 200-0605; Lundström-Åman Foundation; Knut and Alice Wallenberg Foundation, Contracts Nos. 2016.0157, 2021.0299; Swedish Research Council, Contracts Nos. 2019-04594, 2021-04567; Swedish Foundation for International Cooperation in Research and Higher Education, Contract No. CH2018-7756; Istituto Nazionale di Fisica Nucleare, Italy; Ministry of Development of Turkey under Contract No. DPT2006K-120470; National Research Foundation of Korea under Contract No. NRF-2022R1A2C1092335; National Science and Technology fund of Mongolia; National Science Research and Innovation Fund (NSRF) via the Program Management Unit for Human Resources \& Institutional Development, Research and Innovation of Thailand under Contract No. B16F640076; Polish National Science Centre under Contract No. 2019/35/O/ST2/02907; The Swedish Research Council; U. S. Department of Energy under Contract No. DE-FG02-05ER41374.
\input{./Xi0Xi0.bbl}

\end{document}

%% file: authorlist_2023-02-20.tex
{\small M.~Ablikim$^{1}$, M.~N.~Achasov$^{5,b}$, P.~Adlarson$^{75}$, X.~C.~Ai$^{81}$, R.~Aliberti$^{36}$, A.~Amoroso$^{74A,74C}$, M.~R.~An$^{40}$, Q.~An$^{71,58}$, Y.~Bai$^{57}$, O.~Bakina$^{37}$
% ...
\\
\vspace{0.2cm}
(BESIII Collaboration)\\
\vspace{0.2cm} {\it
$^{1}$ Institute of High Energy Physics, Beijing 100049, People's Republic of China\\
% ...
$^{5}$ Budker Institute of Nuclear Physics SB RAS (BINP), Novosibirsk 630090, Russia\\
% ...
$^{36}$ Johannes Gutenberg University of Mainz, Johann-Joachim-Becher-Weg 45, D-55099 Mainz, Germany\\
$^{37}$ Joint Institute for Nuclear Research, 141980 Dubna, Moscow region, Russia\\
% ...
$^{40}$ Liaoning Normal University, Dalian 116029, People's Republic of China\\
% ...
$^{57}$ Southeast University, Nanjing 211100, People's Republic of China\\
$^{58}$ State Key Laboratory of Particle Detection and Electronics, Beijing 100049, Hefei 230026, People's Republic of China\\
% ...
$^{71}$ University of Science and Technology of China, Hefei 230026, People's Republic of China\\
% ...
$^{74}$ University of Turin and INFN, (A)University of Turin, I-10125, Turin, Italy; (B)University of Eastern Piedmont, I-15121, Alessandria, Italy; (C)INFN, I-10125, Turin, Italy\\
$^{75}$ Uppsala University, Box 516, SE-75120 Uppsala, Sweden\\
% ...
$^{81}$ Zhengzhou University, Zhengzhou 450001, People's Republic of China\\
\vspace{0.2cm}
% ...
$^{b}$ Also at the Novosibirsk State University, Novosibirsk, 630090, Russia\\
% ...
}\vspace{0.4cm}  }

%% file: Xi0Xi0.bbl
%